\documentclass[12pt]{article}
\pdfoutput=1
\usepackage{cite,color,url}
\usepackage[colorlinks=true
,urlcolor=blue
,anchorcolor=blue
,citecolor=blue
,filecolor=blue
,linkcolor=blue
,menucolor=blue
,linktocpage=true
,pdfproducer=medialab
,pdfa=true
]{hyperref}

\evensidemargin \oddsidemargin
\newcommand{\customfootnotetext}[2]{{
  \renewcommand{\thefootnote}{#1}
  \footnotetext[0]{#2}}}

\begin{document}

\begin{center}
 
\begin{Large}
\textbf{{On the Quantization of Length in\\ [0.25 em]Noncommutative Spaces}}
\end{Large}
\vspace{1cm}

Muthukumar Balasundaram\textsuperscript{$\dagger$}\customfootnotetext{$\dagger$}{corresponding author}\footnote{{\tt \href{mailto:muthukbs@pondiuni.ac.in}{muthukbs@pondiuni.ac.in}}} and Aamir Rashid\footnote{{\tt \href{mailto:aamirjamian@gmail.com}{aamirjamian@gmail.com}}}%

\vskip 0.6cm

\sl{Department of Physics, \\ School of Physical, Chemical and Applied Sciences,\\ Pondicherry University, Puducherry-605014, India
}
\end{center}

\vspace{1cm}

\begin{abstract}
We consider canonical/Weyl-Moyal type noncommutative (NC) spaces with rectilinear coordinates. Motivated by the analogy of the formalism of the quantum mechanical harmonic oscillator problem in quantum phase-space with that of the canonical-type NC 2-D space, and noting that the square of length in the latter case is analogous to the Hamiltonian in the former case, we arrive at the conclusion that the length and area are quantized in such an NC space, if the area is expressed entirely in terms of length. We extend our analysis to the 3-D case and formulate a ladder operator approach to the  quantization of length in 3-D space. However, our method does not lend itself to the quantization of spacetime length in 1+1 and 2+1 Minkowski spacetimes if the noncommutativity between time and space is considered. If time is taken to commute with spatial coordinates and the noncommutativity is maintained only among the spatial coordinates in 2+1 and 3+1 dimensional spacetime, then the quantization of spatial length is possible in our approach.  
\end{abstract}

\section{Introduction\label{sec-intro}}
Noncommutative (NC) spacetime, which was first introduced by Snyder \cite{Snyder:1946qz} in an attempt to regulate the divergences in quantum field theories, has also been introduced in various contexts \cite{Filk:1996dm,Doplicher:1994tu,Kempf:1994su,Seiberg:1999vs,Madore:1999bi,Nicolini:2008aj} and the literature in this area is quite replete. To name a few, field theories \cite{Szabo:2001kg,Muthukumar:2004wj,Muthukumar:2014hda}, gravity theories \cite{Aschieri:2005yw,Calmet:2005qm,Harikumar:2006xf,Balachandran:2006qg,Roy:2022rav}, and quantum mechanics \cite{Nair:2000ii,Bellucci:2001xp,Muthukumar:2002cn,Muthukumar:2006ab,Muthukumar:2007zza,Biswas:2019obt,Bolonek:2002cc,Muhuri:2020did,Harikumar:2004qc} have all been considered with the background spacetime  being noncommutative. In another development, the spectral manifolds in NC geometry is shown to exhibit the geometric analogue of Heisenberg commutation relation involving the  Dirac operator and the Feynman slash operator for real scalar fields, leading to the quantization of volume \cite{Chamseddine:2014nxa}. In \cite{Roy:2022rav}, it was shown from the pure geometrical point of view that the NC Minkowski spacetime parametrized with spherical or cylindrical coordinates has nontrivial NC corrections to curvature tensors and curvature scalar. 

Noncommutativity of spatial coordinates is  related to the presence of a minimal length in a system. This minimal length in turn is usually related to the uncertainties in the distance measurements \cite{Doplicher:1994tu,Bolonek:2002cc}. Instead of relating the minimal length with uncertainties, we propose in this work an approach in which the actual square of the distance $L^2=g_{ij}(y^i-z^i)(y^j-z^j)$ between any two points $y$ and $z$ in a commutative flat-spacetime is promoted as an operator with the introduction of the canonical/Weyl-Moyal type noncommutativity $[\hat{y}^i,\hat{y}^j]=i\theta'^{ij}=[\hat{z}^i,\hat{z}^j]$ among the coordinate operators. Here $g_{ij}$ is taken to be a constant diagonal metric of spacetime and $\theta'^{ij}$ is a constant and real antisymmetric matrix. The operators $\hat{y}^i$ and $\hat{z}^i$ may be taken either as the position operators of two particles or as the operators corresponding to the positions at which fields are considered. The idea of length as an operator has already been discussed in the literature in the context of canonical quantum gravity \cite{Thiemann:1996at}. We set up an algebra of operators in such a way that the eigenvalues of the operator $\hat{L}^2$ can be raised or lowered. The hint to such an approach is provided by an analogy of 2-D NC space operator formalism with that of the quantum mechanical harmonic oscillator problem. Taking $\hat{x}^i=\hat{y}^i-\hat{z}^i$ and assuming that the coordinate operators of different particles commute, i.e., $[\hat{y}^i,\hat{z}^j]=0$, we can define the operator corresponding to the square of the distance as 
\begin{eqnarray}
\hat{L}^2=g_{ij}\hat{x}^i\hat{x}^j \label{eq-sq-len}
\end{eqnarray}
with 
\begin{eqnarray}
[\hat{x}^i,\hat{x}^j]=i\theta^{ij},  \label{eq-fun-com-rel}
\end{eqnarray}
where $\theta^{ij}=2\theta'^{ij}$. If $g_{ij}=\mathrm{diag}(1,1)$, then the operator $\hat{L}^2$ in such a 2-D NC space can be related to the Hamiltonian of an appropriate harmonic oscillator. One important thing in such an analogy is the set of ladder operators $\hat{a}_-$, the lowering operator, and $\hat{a}_+=\hat{a}_{-}^{\dagger}$, the raising operator. If $\hat{X}=\left(\begin{array}{l}\hat{x}^1\\ \hat{x}^2\end{array}\!\right)$ and $\hat{A}=\left(\begin{array}{l}\hat{a}_-\\ \hat{a}_+\end{array}\!\right)$, then the ladder operators have the following three important relations: 
%
\begin{eqnarray}
[\hat{a}_-,\hat{a}_+]&=&1 \label{eq-lad-com}\\ 
\left[\hat{L}^2,\hat{a}_{\mp}\right]&=&\mp\,\lambda\,\hat{a}_{\mp} \label{eq-len-sq-lad-com}\\
\hat{A}^{\dagger}\hat{A} & =& \gamma \,\hat{X}^{\dagger}\hat{X} =\gamma\,\hat{L}^2, \label{eq-lad-len-rel} 
\end{eqnarray}
%
where $\lambda$ and $\gamma$ are constants.  
If $\hat{A}=U\hat{X}$ denotes the transformation from $\hat{X}$ to $\hat{A}$, then Eq.(\ref{eq-lad-len-rel}) implies that $U^{\dagger}U=\gamma$. With $\theta^{12}=-\theta^{21}=\theta$ in 2-D, we can define the ladder operators as
\begin{eqnarray}
 \hat{a}_{\mp}=\frac{e^{\pm i\delta}}{\sqrt{2\theta}}(\hat{x}^1\pm\,i\,\hat{x}^2) \label{eq-lad-op-2d}
\end{eqnarray}
that satisfy Eq.(\ref{eq-lad-com})-Eq.(\ref{eq-lad-len-rel}) with $\lambda=2\theta$ and $\gamma=1/\theta$. Here $\delta$ is an arbitrary real number. The square of the length in this case is given by
\begin{eqnarray}
 \hat{L}^2=2\theta\left(\hat{a}_+\hat{a}_- +\frac{1}{2}\right). \label{eq-sq-len-2d}
\end{eqnarray}
In analogy with the quantum harmonic oscillator problem, it is clear that on the eigenstate $|n\rangle$ of $\hat{a}_+\hat{a}_-$,
\begin{eqnarray}
 \hat{L}^2\,|n\rangle=2\theta[n+(1/2)]\, |n\rangle, \label{eq-len-sq-quanta}
\end{eqnarray}
with the minimum eigenvalue being $\theta$. The operator $\hat{a}_+$ plays the role of  a raising  operation i.e., the eigenvalues of  $\hat{L}^2$ on the states $\left|n\right>$ and $\hat{a}_+\!\left|n\right>$ are $2\theta$ apart.  The operator $\hat{a}_-$ does the lowering operation. The angle $\delta$ in Eq.(\ref{eq-lad-op-2d}) \emph{corresponds} to the orientation of the line segment that would connect the points $y$ and $z$ in the commutative 2-D space.

On such admissible eigenstates, the square-root of Eq.(\ref{eq-sq-len-2d}) is also quantized. In 2-D Euclidean space, the commutative analogue of Eq.(\ref{eq-sq-len}) could also mean the area of a square of side $L$ in which case the quantization of Eq.(\ref{eq-sq-len-2d}) can also imply the quantization of area. In \cite{Chamseddine:2014nxa}, it has been shown in the context of spectral manifolds in NC geometry that the area of a 2-D manifold is quantized. Here, we arrive at the same result in the case of 2-D plane with coordinate noncommutativity. In the following, we attempt to generalize this result to other NC spaces. 

Before proceeding, note that the commutator Eq.(\ref{eq-len-sq-lad-com}) is essential to construct a ladder of states which are all eigenstates of $\hat{L}^2$ operator and that the canonical commutator Eq.(\ref{eq-fun-com-rel}) (or Eq.(\ref{eq-lad-com})) can be used to reduce the degree of any polynomial $\hat{P}(\hat{x})$ by considering the commutator $[\hat{P}(\hat{x}),\hat{x}^i]$.  Eq.(\ref{eq-len-sq-lad-com}) is an example for such reduction of the degree of a polynomial. In this context, it is worth remarking that a generalized coordinate noncommutativity, which would involve a non-constant $\theta^{ij}$ in Eq.(\ref{eq-fun-com-rel}), will pose problem to construct a ladder of eigenstates of $\hat{L}^2$ using Eq.(\ref{eq-len-sq-lad-com}). So such generalized coordinate commutator structure (see for example \cite{Kempf:1994su}) is not considered in this work\footnote{The authors thank the referee for raising the issue of generalized coordinate commutator structure.}. 

\section{A Generalization}
It turns out that Eq.(\ref{eq-lad-com})-Eq.(\ref{eq-lad-len-rel}) play a crucial role in the generalization of the idea of length quantization to other spaces with constant and symmetric metric $g_{ij}$. In an $N$-dimensional space we define the lowering operator and the operator corresponding to square of length respectively as
\begin{eqnarray}
 \hat{a}_-= \alpha_i \, \hat{x}^{i}, \qquad \hat{L}^2=g_{ij}\hat{x}^i\hat{x}^j, \qquad i=1,2,\ldots N, \label{eq-lad-op-exp-n-d}
\end{eqnarray}
where Einstein's summation convention is implied and $\alpha_i$'s are complex constants to be determined. Substitution of Eq.(\ref{eq-lad-op-exp-n-d}) in Eq.(\ref{eq-len-sq-lad-com}) gives the relation
\begin{eqnarray}
 2i\,g_{ij}\theta^{jk}\alpha_k = -\lambda \alpha_i. \label{eq-alpha-n-d}
\end{eqnarray}
Also, Eq.(\ref{eq-lad-com}) and Eq.(\ref{eq-alpha-n-d}) leads to 
\begin{eqnarray}
 \alpha^i \alpha_i^*  = 2/\lambda, \label{eq-alpha-alpha-star}
\end{eqnarray}
where $\alpha_i^*$ is the complex conjugate of $\alpha_i$. 
In the following, we solve Eq.(\ref{eq-alpha-n-d}) and Eq.(\ref{eq-alpha-alpha-star}) for the cases of 3-D space, 1+1 dimensional spacetime and 2+1 dimensional spacetime, and analyze the consequences. 
\subsection{The Case of 3-D}
We assume that $g_{ij}=\mathrm{diag}(1,1,1)$ and that $\theta^{12}=\theta^{13}=\theta^{23}=\theta$, in which case the set of three equations Eq.(\ref{eq-alpha-n-d}) have nontrivial solution only if the secular determinant
\begin{eqnarray}
\left|
 \begin{array}{ccc}
 \lambda & 2i\theta & 2i\theta \\
 -2i\theta & \lambda & 2i\theta \\
 -2i\theta & -2i\theta & \lambda
 \end{array}
\right| =0, 
\end{eqnarray}
which leads to the nontrivial value $\lambda=\pm 2\sqrt{3}\,\theta$. The third trivial solution $\lambda=0$ leads to a set of $\alpha_i$'s such that $\hat{a}_+ \propto \hat{a}_-$, thereby violating Eq.(\ref{eq-lad-com}). Also, $\lambda=0$ does not give any length-raising or lowering operation in the theory (see Eq.(\ref{eq-len-sq-lad-com})). Choosing the positive value for $\lambda$ so that $\hat{a}_-$ can be identified with a lowering operation and substituting it in Eq.(\ref{eq-alpha-n-d}) and Eq.(\ref{eq-alpha-alpha-star}) gives a  set of values for $\alpha_i$'s: 
\begin{eqnarray}
 (\alpha_1,\,\alpha_2,\,\alpha_3)= 
 \left(\rho\sigma, \, -\rho\sigma^*, \, \rho\right), \label{eq-alphas-3d}
\end{eqnarray}
where $\rho= \displaystyle e^{i\delta_1}/(\sqrt{3\theta\sqrt{3}})$ and $\sigma=-e^{i\pi/3}$, and $\delta_1$ is a real constant. Note that  $|\alpha_i|=|\rho|=1/(\sqrt{3\theta\sqrt{3}})$. The lowering operator can then be expressed as 
\begin{eqnarray}
 \hat{a}_-=\rho \left[\sigma \hat{x}^1- \sigma^* \hat{x}^2+\hat{x}^3\right].  \label{eq-lower-op-3d}
\end{eqnarray}

Since there are three independent operator variables $\hat{x}^1,\,\hat{x}^2$ and $\hat{x}^3$, we need another operator $\hat{b}$ in addition to $\hat{a}_-$ and $\hat{a}_+$, to define the transformation $\hat{A}=U\hat{X}$ in 3-D and to make it invertible. But $\hat{b}$ and its Hermitian conjugate $\hat{b}^{\dagger}$ should not be independent of each other, else we would end up with four independent operator variables. We define  in the following way:
\begin{eqnarray}
 \hat{b}=\beta_i\,\hat{x}^i, \label{eq-b-expres}
\end{eqnarray}
where $\beta_i$'s are complex numbers. Since $\hat{b}$ is defined as a linear and homogeneous function of $\hat{x}^i$, its Hermitian conjugate $\hat{b}^\dagger$ is also linear and homogeneous in $\hat{x}^i$ and so $\beta_i^*=\beta\beta_i$, where $\beta$ is such that $|\beta|=1$ since $|\beta_i^*|=|\beta_i|$. If $\hat{A}^{\dagger}=(\hat{a}_+,\, \hat{a}_-, \, \hat{b}^{\dagger})$ and $\hat{X}^{\dagger}=(\hat{x}^1,\,\hat{x}^2, \, \hat{x}^3)$, then  the analog of Eq.(\ref{eq-lad-len-rel}) is written as
\begin{eqnarray}
 \hat{L}^2  = \hat{X}^\dagger g \hat{X}= \frac{1}{\gamma}\hat{A}^\dagger g\,\hat{A}=\frac{1}{\gamma}\hat{X}^\dagger U^\dagger g \,U \hat{X},\nonumber
 \end{eqnarray}
where $g$ is the matrix form of the metric tensor. The above $\hat{L}^2$ can further be expressed as
 \begin{eqnarray}
\hat{L}^2 = (\hat{x}^1)^2+(\hat{x}^2)^2+(\hat{x}^3)^2   =\frac{1}{\gamma}(2\hat{a}_+\hat{a}_-+\hat{b}^{\dagger}\hat{b}+1)
 \label{eq-len-sq-exp-3d}
\end{eqnarray}
upon the condition that $U^\dagger g\, U=\gamma\, g$. Using the expressions for $\hat{a}_-$, $\hat{a}_+$ and $\hat{b}$ in terms of $\hat{x}^i$ as in Eq.(\ref{eq-lad-op-exp-n-d}) and Eq.(\ref{eq-b-expres}), we can get the elements of $U$ in terms of $\alpha_i$, $\alpha_i^*$ and $\beta_i$ from which the condition $U^\dagger g\, U=\gamma\, g$ in indicial form is expressed as 
\begin{eqnarray}
\alpha_i^*\alpha_j+\alpha_i\alpha_j^*+\beta_i^*\beta_j=\gamma \, g_{ij}. \label{eq-alpha-beta-rel}
\end{eqnarray}
For all the values $i=j$, the above condition gives $|\beta_j|$ as 
\begin{eqnarray}
 |\beta_j|=\sqrt{(\gamma-2|\rho|^2)}. \label{eq-beta-val1}
\end{eqnarray}
For $i=1$ and $j=2$ and $3$ we can respectively show that 
\begin{eqnarray}
 \beta_2 =-\frac{1}{\beta_1^*}(\alpha_1^*\alpha_2+\alpha_1\alpha_2^*),\qquad
 \beta_3 =-\frac{1}{\beta_1^*}(\alpha_1^*\alpha_3+\alpha_1\alpha_3^*),
\label{eq-beta-val2}
 \end{eqnarray}
which in general leads to 
%
$\beta_i^*=\beta_i(\beta_1^*/\beta_1)$. Calculation of $|\beta_1^*\beta_2|$ using Eq.(\ref{eq-beta-val2}), (\ref{eq-beta-val1}) and (\ref{eq-alphas-3d}) gives $\gamma=3|\rho|^2=1/(\sqrt{3}\,\theta)$. Therefore, by writing $\beta_1=|\rho|\, e^{i\delta_2}$, Eq.(\ref{eq-beta-val2}) completely determines $\beta_2$ and $\beta_3$ in terms of $\delta_2$ and $|\rho|$:
\begin{eqnarray}
 (\beta_1,\,\beta_2,\,\beta_3)=(|\rho|e^{i\delta_2},\,-|\rho|e^{i\delta_2},\, |\rho|e^{i\delta_2}). \label{eq-beta-val3}
\end{eqnarray}
Then the only free parameters in the theory would be $\delta_1$ and $\delta_2$. These two parameters would correspond to the orientation of the line segment connecting the points $y$ and $z$ in the physical 3-D commutative space. 

Note that the explicit values Eq.(\ref{eq-alphas-3d}) of the $\alpha_i$'s and $\beta_i$'s have the following properties:
\begin{eqnarray}
 \alpha_i\,\alpha^i&=&0, \label{eq-alpha-alpha}\\ 
 \theta^{ij}\beta_j& =&0. \label{eq-theta-beta} \\
 \alpha_i\,\beta^i&=&\alpha_i^{*}\,\beta^i=0, \label{eq-alpha-beta}
\end{eqnarray}
Eq.(\ref{eq-alpha-alpha}) can also be inferred from the $[\hat{a}_-,\hat{a}_-]=0$. Eq.(\ref{eq-alpha-beta}) leads to $[\hat{a}_-,\hat{b}]=[\hat{a}_+,\hat{b}]=0$. In fact, Eq.(\ref{eq-theta-beta}) leads to $[\hat{x}^i,\hat{b}]=0$ and thus $\hat{b}$ commutes with all the operators of the form $\hat{F}(\hat{x})$ in the theory. In other words, the states characterized by any index corresponding to $\hat{b}$ are unaffected by other operators of the form $\hat{F}(\hat{x})$.

Using Eq.(\ref{eq-len-sq-lad-com}), it is easy to show that $[\hat{L}^2, \hat{a}_+\hat{a}_-]=0$. So it is possible to construct a complete set of simultaneous eigenstates of $\hat{L}^2$, $\hat{a}_+\hat{a}_-$ and $\hat{b}^{\dagger}\hat{b}$.

Like in the case of quantum harmonic oscillator problem, the relations Eq.(\ref{eq-lad-com}) and Eq.(\ref{eq-len-sq-lad-com}) ensure that the eigenvalues of $\hat{a}_+\hat{a}_-$ should be non-negative integers (since the eigenvalues of $\hat{b}^{\dagger}\hat{b}$ will also turn out to be non-negative), and the eigenstate itself may be denoted by $|n\rangle$. Using the commutator 
\begin{eqnarray}
[\hat{x}^i,\,\hat{a}_+\hat{a}_-]= \frac{\lambda}{2}( \alpha^{i*}\,\hat{a}_--\alpha^i\,\hat{a}^{\dagger}), \label{eq-bdaggerb-lad-com-rel}
\end{eqnarray}
we can show that
\begin{eqnarray}
 \hat{a}_+\hat{a}_-(\hat{x}^i|n\rangle)=-\frac{\lambda}{2}\left( \alpha^{i*}\,\sqrt{n}|n-1\rangle -\alpha^i\,\sqrt{n+1}|n+1\rangle \right) + n (\hat{x}^i|n\rangle). \label{eq-no-op-on-xin}
\end{eqnarray}
The above relation suggests that the state $(\hat{x}^i|n\rangle)$ can be written as 
\begin{eqnarray}
 \hat{x}^i|n\rangle=\chi_{(n)-1}^i|n-1\rangle+\chi_{(n)}^i|n\rangle+\chi_{(n)+1}^i|n+1\rangle, \label{eq-xin}
\end{eqnarray}
where $\chi_{(n)-1}^i,~\chi_{(n)}^i$ and $\chi_{(n)+1}^i$ are constants to be determined. 

By acting $\hat{a}_+\hat{a}_-$ on Eq.(\ref{eq-xin}) and comparing the result with Eq.(\ref{eq-no-op-on-xin}), we have 
\begin{eqnarray}
 \chi_{(n)-1}^i=\frac{\lambda}{2}\alpha^{i*}\sqrt{n}\,, \qquad \chi_{(n)+1}^i=\frac{\lambda}{2}\alpha^{i}\sqrt{n+1} \label{eq-xin-pm-one}
\end{eqnarray}
Projecting the state Eq.(\ref{eq-xin}) onto $|n\rangle$, $|n-1\rangle$ and $|n+1\rangle$ respectively and in those projections comparing the action of $\hat{x}^i$ on the ket vector with its action on the bra vector lead to 
\begin{equation}
 \chi_{(n)}^{i*} =\chi_{(n)}^i, \label{eq-chin-reality} \qquad
 \chi_{(n-1)+1}^{i*}=\chi_{(n)-1}^i, \qquad 
 \chi_{(n+1)-1}^{i*} =\chi_{(n)+1}^i.
\end{equation}

Making use of the commutator $[\hat{x}^i,\,\hat{a}^{\dagger}]=\frac{\lambda}{2}\alpha^{i*}$ in the calculation of $\langle n+1|a^{\dagger}\hat{x}^i|n\rangle$, we can show that $\chi_{(n+1)}^i=\chi_{(n)}^i$ and therefore $\chi_{(n)}^i$ is independent of $n$, which we denote by $\chi^i$. To be consistent with the relations like $\hat{a}|n\rangle=\sqrt{n}|n-1\rangle$, $\chi^i$ needs to satisfy
\begin{eqnarray}
\chi^i\,\alpha_i=0=\chi^i\alpha_i^*, \label{eq-chi-alpha}
\end{eqnarray}
which is possible 
if we choose anyone of the following set for $\chi^i$:
\begin{eqnarray}
 \chi^i= (+\chi, \, -\chi, \, +\chi) ~~ \mathrm{or} ~~
                      (-\chi, \, +\chi,\, -\chi),
\label{eq-chii-values}
 \end{eqnarray}
which gives $(\chi^i)^2=\chi^2$ for any particular $i$, where $\chi$ is some real constant. The values in Eq.(\ref{eq-chii-values}) are motivated by the values of $\beta^i$, Eq.(\ref{eq-beta-val3}), because both $\chi^i$ and $\beta^i$ have the same properties with $\alpha_i$ (compare Eq.(\ref{eq-chi-alpha}) and Eq.(\ref{eq-alpha-beta})). 

If we choose the states $\left(\frac{1}{\sqrt{\theta\xi^i_n}}\hat{x}^i|n\rangle\right)$ to normalize to $1$, where $i$ is not summed over and $\xi^i_n$'s are dimensionless constants to be determined, then using Eq.(\ref{eq-xin}) and its Hermitian conjugate,  the normalization leads to
\begin{eqnarray}
 |\xi_n^i|=\frac{1}{\theta}\left(\chi^2+\frac{\lambda^2|\rho|^2}{4}(2n+1)\right). \label{eq-xin-value}
\end{eqnarray}
We denote the above expression by $\xi_n(\chi)$ since  it is the same for all $i$ and it depends on $\chi$. Also, using Eq.(\ref{eq-alpha-beta}), Eq.(\ref{eq-xin-pm-one}) and Eq.(\ref{eq-chii-values}), it is straightforward to show that 
\begin{eqnarray}
\hat{b}|n\rangle &=& (\beta_i\chi^i)|n\rangle= (3|\rho|\chi e^{i\delta_2}) |n\rangle, \\
 \hat{b}^{\dagger}\hat{b}|n\rangle &=&(\beta_i^*\,\chi^i)(\beta_j\,\chi^j)|n\rangle=(\frac{\sqrt{3}\,\chi^2}{\theta})|n\rangle,
\end{eqnarray}
and therefore the states may be properly denoted by $|n,\chi\rangle$ instead of $|n\rangle$. Since the operator $\hat{b}$ commutes with all other operators, it is not possible to determine $\chi$ through the  operator algebra method. 

Finally, the eigenvalues of $\hat{L}^2$ in Eq.(\ref{eq-len-sq-exp-3d}) are worked out to be
\begin{eqnarray}
 \hat{L}^2|n,\chi\rangle=\theta\, \xi_n\,|n,\chi\rangle=(\chi^2+\frac{\theta}{\sqrt{3}}(2n+1))|n,\chi\rangle. 
\end{eqnarray}
Essentially, we have worked out the eigenvalues of $\hat{L}^2$ in terms of the eigenvalues corresponding to $\hat{a}_+\hat{a}_-$ and $\hat{b}^{\dagger}\hat{b}$. Since $\chi$ is real because of Eq.(\ref{eq-chin-reality}), the minimum value for $\chi^2$ is $0$, and therefore we have, 
\begin{eqnarray}
 \hat{L}^2|n,0\rangle=\theta\,\xi_n\,|n,0\rangle=\frac{\theta}{\sqrt{3}}(2n+1)|n,0\rangle. 
\end{eqnarray}
In the Euclidean 3-D space, the commutative analogue of Eq.(\ref{eq-len-sq-exp-3d}) can also be interpreted as the area of a square of side $L$ in the plane formed by $\hat{a}_-$ and $\hat{a}_+$, and so the quantization of Eq.(\ref{eq-len-sq-exp-3d}) implies the quantization of area with the minimum value being $\frac{\theta}{\sqrt{3}}=\frac{2\theta'}{\sqrt{3}}$. But the actual uncertainty relation $(\Delta y^1)(\Delta y^2)\geq\frac{\theta'}{2}$ would yield the minimum $\frac{\theta'}{2}$ which is lower than $\frac{2\theta'}{\sqrt{3}}$. So the minimum value of the  quantized area is not violating the uncertainty principle. If a volume is written as a function of the side $\hat{L}$, then the volume is also quantized along the dimensions of $\hat{a}_-$ and $\hat{a}_+$.

\subsection{1+1 Dimensional Spacetime}
If $g_{ij}=\mathrm{diag}(1,-1)$ and $\theta^{12}=-\theta^{21}=\theta$, then the relation Eq.(\ref{eq-alpha-n-d}) leads to the purely imaginary value for $\lambda=\pm 2 i\theta $. Also, the substitution of this value back into Eq.(\ref{eq-alpha-n-d}) implies that $\alpha_1=\mp \alpha_2$, which will contradict Eq.(\ref{eq-lad-com}). Therefore, the quantization of spacetime length in this approach is not feasible. 
\subsection{2+1 Dimensional Spacetime}
If we take $g_{ij}=\mathrm{diag}(1,1,-1)$ and $\theta^{12}=\theta^{13}=\theta^{23}=\theta$, then the solution to the secular equation corresponding to Eq.(\ref{eq-alpha-n-d}) gives only the purely imaginary number $\lambda=\pm 2i\theta$ as the nontrivial value. Putting the value $\lambda=2i\theta$ in Eq.(\ref{eq-alpha-n-d}) gives no nontrivial solution to $\alpha_i$, and the value $\lambda=-2i\theta$ results in $\hat{a}^{\dagger} \propto \hat{a}$, contradicting Eq.(\ref{eq-lad-com}). So, in this case also, our method is not feasible to quantize the spacetime length. However, if $\theta^{23}=0$, i.e., if the time commutes with the spatial coordinates, then the method outlined in the Introduction can be used to quantize the spatial part of $\hat{L}^2$ in a particular rectilinear system of coordinates.

\subsection{3+1 Dimensional Spacetime}
In this case, if the time is taken to commute with the spatial coordinates, then the same method for the 3-D case can be used to quantize the spatial length in a particular rectilinear system of coordinates. However, if time is assumed to noncommute with spatial coordinates, it requires a separate elaborate treatment and  it will be published elsewhere
.
\section{Concluding Remarks}
In this work, we have proposed a length operator $\hat{L}$ in Weyl-Moyal type noncommutative (NC) spaces, and analyzed the consequences on the quantization of length. By comparing the operator formalism of 2-D NC space with that of quantum harmonic oscillator problem, we have deduced that the operator corresponding to the square of length is analogous to the Hamiltonian of the oscillator and hence the length and area are quantized in 2-D NC space. This result is in conformity with and a special case of the already established result in the context of spectral manifolds in NC geometry \cite{Chamseddine:2014nxa}. We have also succeeded in showing that the length is quantized in a 2-D NC subspace of a 3-D NC space. Since the length quantization is more fundamental, the quantization of area and volume \cite{Chamseddine:2014nxa} can be inferred from it for the special cases in which they directly depend on the length along the 2-D subspace. But our method does not work for the cases of 1+1 and 2+1 spacetime dimensions if the time is taken to noncommute with the spatial coordinates. This is because the eigenvalue equation Eq.(\ref{eq-alpha-n-d}) which is the result of the commutator Eq.(\ref{eq-len-sq-lad-com}) gives imaginary eigenvalues for $\lambda$ --- the quantum of $\hat{L}^2$. This essentially implies that the length is increased or decreased in steps of imaginary values which is unphysical. When the metric is changed to an Euclidean metric, the quanta of $\hat{L}^2$ become real.

\section*{Data Availability}

No new data were created or analysed in this study.

\section*{Conflicts of Interests}

The authors declare that there is no conflict of interest regarding the publication of this work. 

\section*{Funding Statement}

The authors gratefully acknowledge the funding (No.PU/PS2/PHYS/Minor-Equip/21-22/241) by Pondicherry University under the Minor Equipment Grant No.PU/PD2/Minor-Equip/2022/536.

\section*{Acknowledgment}

A version of this work has been published as an arXiv preprint: https://arxiv.org/abs/2206.07972 \cite{Muthukumar:2022gnk}.

\end{document}